\documentclass[11pt]{article}
\usepackage{moriond,epsfig}

\def\Journal#1#2#3#4{{#1} {\bf #2}, #3 (#4)}

\def\PLB{{\em Phys. Lett.}  B}

\def\PRD{{\em Phys. Rev.} D}

\def\mco{\multicolumn}

\def\ra{\rightarrow}

\def\ko{K^0}

\def\be{\begin{equation}}
\def\ee{\end{equation}}
\def\bea{\begin{eqnarray}}
\def\eea{\end{eqnarray}}

\newcommand{\gsim}{\mbox{\raisebox{-.9ex}{~$\stackrel{\mbox{$>$}}{\sim}$~}}}

\begin{document}
\vspace*{4cm}
\title{MODULAR HYBRID INFLATION 
WITHOUT FLAT DIRECTIONS\\ AND THE CURVATON}

\author{
K. DIMOPOULOS}

\address{
Institute of Nuclear Physics, National Center for Scientific Research
`Demokritos',\\ 
Agia Paraskevi Attikis, Athens 153 10, Greece
}

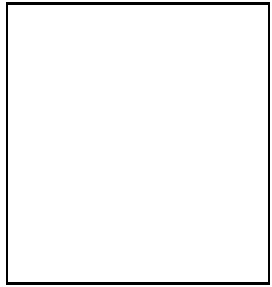
\maketitle\abstracts{
I study two-stage hybrid inflation driven by moduli fields, corresponding to 
flat directions of supersymmetry, lifted by supergravity corrections. The 
first stage corresponds to a period of either fast-roll or `locked' inflation, 
induced by an oscillating inflaton. This is followed by a second stage of 
fast-roll inflation. Enough total e-foldings to encompass the cosmological 
scales are achieved. Structure in the Universe is generated due to a curvaton 
field.
}

\section{Introduction}

The latest CMB observations suggest that structure formation in the Universe
is due to the existence of a {\em superhorizon} spectrum of curvature
perturbations. This strongly implies that, during the early stages of its 
evolution, the Universe underwent a period of cosmic inflation.

According to the inflationary paradigm, inflation is realized through the 
domination of the Universe by the potential density of a light scalar field, 
which is slowly rolling down its almost flat potential. One of the reasons for 
using a flat potential is that one requires inflation to last long enough for 
the cosmological scales to exit the horizon during the period of accelerated 
expansion, so as to solve the horizon and flatness problems. Thus, inflation 
seems to require the presence of a suitable flat direction in field space. 
Unfortunately, this is hard to attain in supergravity because K\"{a}hler 
corrections generically lift the flatness of the scalar potential 
\cite{randall}.

Still, there have been attempts to overcome this so-called $\eta$-problem of 
inflation. A first step toward inflation without a flat direction is 
fast-roll inflation \cite{fastroll}, which, however, may last only for a 
limited number of e-foldings and, hence, it is probably not capable to explain 
the observations. Recently, another mechanism for inflation without a flat 
direction was suggested~\cite{dvali}. Rapid oscillations in a hybrid-type 
potential keep the field `locked' on top of a saddle point and prevent it from 
rolling toward the minima. Unfortunately, oscillatory inflation is also too 
brief.

In this paper 
\cite{ours}
we point out that in a hybrid-type non-flat potential one can 
have two consecutive stages of inflation. Depending on the curvature of the 
potential, the first stage is a period of either fast-roll or oscillatory 
`locked' inflation. This is followed by a second period of tachyonic fast-roll 
inflation. In total, inflation may last long enough to solve the horizon and 
flatness problems, without imposing stringent bounds on the curvature of the 
potential. 
Since our inflaton is not a light field it cannot be responsible for the
observed spectrum of curvature perturbations. We, therefore, consider that 
these perturbations are due to a curvaton field~\cite{curvaton}.

\section{Fast--Roll versus Locked Inflation}

Consider two moduli fields, which parameterize flat directions of supersymmetry
(whose flatness is lifted by supergravity corrections) with a hybrid type of 
potential of the form
\begin{eqnarray}
V(\Phi,\phi)\;=
&
\!\!\frac{1}{2}\,m_\Phi^2\Phi^2+
\frac{1}{2}\,\lambda\Phi^2\phi^2+\frac{1}{4}\,\alpha(\phi^2-M^2)^2 
&,
\label{V}
\end{eqnarray}
where $\Phi$ and $\phi$ above are taken to be real scalar fields with
\mbox{$m_\Phi\sim M_S^2/m_P\sim m_{3/2}$}, \mbox{$M\sim m_P$},
\mbox{$\alpha\sim(M_S/m_P)^4$}, \mbox{$\lambda\sim 1$} and
\mbox{$M_S\sim\sqrt{m_{3/2}m_P}\sim 10^{11}$GeV} 
corresponding to gravity mediated supersymmetry breaking, where
\mbox{$m_{3/2}\sim$ TeV}. The tachyonic mass of $\phi$ is 
\mbox{$m_\phi\sim\sqrt{\alpha}\,M\sim m_{3/2}$}.

The above potential has global minima at \mbox{$(\Phi,\phi)=(0,\pm M)$} and
an unstable saddle point at \mbox{$(\Phi,\phi)=(0,0)$} similarly to hybrid 
inflation \cite{hybrid}. However, in contrast to regular hybrid inflation, for 
\mbox{$|\Phi|,|\phi|\leq m_P$}, the potential does not satisfy the slow-roll
requirements. 

Now, since the effective mass--squared of $\phi$ is
\mbox{$(m_\phi^{\rm eff})^2=\lambda\Phi^2-\alpha M^2$},
$\phi$ is driven to zero if 
\mbox{$\Phi>\Phi_c\equiv\sqrt{\alpha/\lambda}\;M\sim m_{3/2}$}.
Suppose, that originally the system lies in the regime, where,
\mbox{$m_{3/2}<\Phi\leq m_P$} and \mbox{$\phi\simeq 0$}. With such initial 
conditions the effective potential for $\Phi$ becomes:
\mbox{$V(\Phi,\phi=0)=\frac{1}{2}m_\Phi^2\Phi^2+M_S^4$}. Since 
\mbox{$\Phi<m_P$}, when $\phi$ remains at the origin, the scalar potential is 
dominated by a false vacuum density \mbox{$V_{\rm inf}\simeq M_S^4$}, resulting
in a period of inflation. During this period, we have 
\mbox{$V_{\rm inf}\sim(m_PH_{\rm inf})^2$}, which means that 
\mbox{$H_{\rm inf}\sim m_{3/2}$}. This is why there is no slow roll; because 
all the masses are of the order of the Hubble parameter, as expected by the 
action of supergravity corrections \cite{randall}. And yet, there is inflation 
as long as $\phi$ remains at the origin. 

During this period the Klein-Gordon equation for $\Phi$ is:
\mbox{$\ddot{\Phi}+3H_{\rm inf}\dot{\Phi}+m_\Phi^2\Phi=0$}, which
has a solution of the form \mbox{$\Phi\propto e^{\omega t}$},
where \mbox{$\omega=-\frac{3}{2}H_{\rm inf}\left[
1\pm\sqrt{1-\frac{4}{9}(m_\Phi/H_{\rm inf})^2}\;\right]$}.
Therefore, the evolution of $\Phi$ depends on whether $m_\Phi$ is larger or 
not from $\frac{3}{2}H_{\rm inf}$.

\subsection{Fast--Roll Inflation ($m_\Phi\leq\frac{3}{2}H_{\rm inf}$)}

In this case, there are two exponential solutions to the Klein-Gordon 
equation, both decreasing with time. The solution with the positive sign 
decreases faster and rapidly disappears. Thus, 
\begin{eqnarray}
\Phi=\Phi_0\exp(-F_\Phi\Delta N)\,, & 
\quad
{\rm with}\qquad
F_\Phi\equiv\frac{3}{2}\left[1-\sqrt{1-\frac{4}{9}
(m_\Phi/H_{\rm inf})^2
}\;\right] &
,
\label{PhiN&F}
\end{eqnarray}
where, \mbox{$\Delta N\equiv H_{\rm inf}\Delta t$}. 
%
Therefore, the total number of e-foldings of fast-roll inflation is 
\begin{eqnarray}
N_{\rm FR}\;= &
\!\!-\frac{1}{F_\Phi}\,\ln(\Phi_{\rm end}/\Phi_0)
\simeq \frac{1}{F_\Phi}\,\ln(m_P/m_{3/2}) & ,
\label{Nfr1}
\end{eqnarray}
where \mbox{$\Phi_0\sim m_P$} and \mbox{$\Phi_{\rm end}=\Phi_c\sim m_{3/2}$}. 
The larger $m_\Phi$ is the smaller is the number $N_{\rm FR}$ of the total 
e-foldings of fast-roll inflation. However, this number cannot become 
arbitrarily small because, if $m_\Phi$ is bigger than 
$\frac{3}{2}H_{\rm inf}$, then the dynamics of $\Phi$ becomes distinctly 
different.

\subsection{Locked Inflation ($m_\Phi>\frac{3}{2}H_{\rm inf}$)}

In this case the Klein-Gordon equation for $\Phi$ is solved by an equation of 
the form:
\begin{eqnarray}
\Phi=
{\Phi}_0\,
e^{-\frac{3}{2}\,\Delta N}
\cos(\omega_\Phi\Delta t)\,, & 
\quad{\rm with}\qquad
\omega_\Phi=H_{\rm inf}
\sqrt{(m_\Phi/H_{\rm inf})^2-\frac{9}{4}}
\;\approx m_\Phi
&.
\label{Phiosc}
\end{eqnarray}
The field is oscillating instead of rolling toward the true minimum of the 
potential because, provided the frequency of the oscillations is large enough, 
the time $(\Delta t)_s$ that the system spends on top of the saddle point 
is too small to allow its escape from the oscillatory trajectory. Indeed, 
\mbox{$(\Delta t)_s\sim\Phi_c/m_\Phi\bar{\Phi}\sim\bar{\Phi}^{-1}$},
where $\bar{\Phi}$ is the amplitude of the oscillations. Originally 
\mbox{$\bar{\Phi}\sim m_P$} but the expansion of the Universe dilutes the 
energy of the oscillations and $\bar{\Phi}$ decreases, which means that 
$(\Delta t)_s$ grows. However, until $(\Delta t)_s$ becomes large enough to be 
comparable to $m_\phi^{-1}$, $\phi$ has no time to roll away from the saddle. 
Hence, the oscillations continue until 
\mbox{$\bar{\Phi}_{\rm end}\sim m_{3/2}\sim\Phi_c$},
at which point $\phi$ departs from the origin and rolls down toward its VEV.

During the oscillations the density of the oscillating $\Phi$ is 
\mbox{$\rho_\Phi=\frac{1}{2}\dot{\Phi}^2+\frac{1}{2}m_\Phi^2\Phi^2\simeq
\frac{1}{2}m_\Phi^2\bar{\Phi}^2$}. Hence, for \mbox{$\bar{\Phi}<m_P$}, 
the overall density is dominated by $V_{\rm inf}$, which remains constant as 
long as $\phi$ remains locked at the origin. Consequently, the Universe 
undergoes inflation when $\bar{\Phi}$ lies in the range 
\mbox{$\bar{\Phi}\in (m_{3/2},m_P)$}. Therefore, the total number of
e-foldings of locked inflation is 
\begin{eqnarray}
N_{\rm lock}\;=&
\!\!\frac{2}{3}\,\ln(m_P/m_{3/2})\simeq 24
&.
\label{Nlock}
\end{eqnarray}
Hence, \mbox{$N_{\rm FR}>N_{\rm lock}$}, i.e. {\em $N_{\rm lock}$ is the 
minimum number of e-foldings that the Universe inflates while $\phi$ remains 
at the origin}. Thus, regardless of the curvature along the $\Phi$-direction,
a minimum number of e-foldings is guaranteed. However, locked inflation alone 
cannot provide the necessary number of e-foldings corresponding to the 
cosmological scales. Fortunately, there is a subsequent period of inflation, 
driven by the scalar field $\phi$ after it departs from the origin.

\section{Tachyonic Fast--Roll Inflation}

The potential for $\phi$ is: \mbox{$V(\phi)=V_{\rm inf}-\frac{1}{2}
|(m_\phi^{\rm eff})^2|\phi^2+\frac{1}{4}\alpha\phi^4$}. 
Since the roll of $\phi$ begins after \mbox{$\bar{\Phi}<\Phi_c$}, we have
\mbox{$|(m_\phi^{\rm eff})^2|\sim m_\phi^2$}. The Klein-Gordon satisfied by 
$\phi$ is: \mbox{$\ddot{\phi}+3H_{\rm inf}\dot{\phi}-m_\phi^2\phi=0$},
which admits solutions of the form \mbox{$\phi\propto e^{\omega_\phi t}$} 
with \mbox{$\omega_\phi=-\frac{3}{2}\;H_{\rm inf}
\left[1\pm\sqrt{1+\frac{4}{9}(m_\phi/H_{\rm inf})^2}\;\right]$}.
The positive sign solution corresponds to the decreasing mode which rapidly 
disappears. Hence,
\begin{eqnarray}
\phi=\phi_0\exp(F_\phi\Delta N)\,, &
\quad{\rm with}\qquad
F_\phi\equiv\frac{3}{2}\left[\sqrt{1+\frac{4}{9}
(m_\phi/H_{\rm inf})^2
}-1\right]
&.
\label{phiN&F}
\end{eqnarray}

From the above, we see that the total number of e-foldings of tachyonic 
fast-roll inflation is 
\begin{eqnarray}
N_\phi\;=&
\!\!\frac{1}{F_\phi}\,\ln(\phi_{\rm end}/\phi_0)
\simeq \frac{1}{F_\phi}\,\ln(M/m_\phi)
&,
\label{Nfr2}
\end{eqnarray}
where the final value of $\phi$ is its VEV: \mbox{$M\sim m_P$}, while the 
initial value of $\phi$ is determined by the tachyonic fluctuations which send 
it off the top of the potential, and is given by \cite{felder}
\mbox{$\phi_0=m_\phi/2\pi$}.

From Eqs.~(\ref{Nfr1}), (\ref{Nlock}) and (\ref{Nfr2}) we see that the total 
number of inflationary e-foldings is 
\begin{eqnarray}
N_{\rm tot}=N_\Phi+N_\phi\;\simeq & 
\!\!\left(\frac{1}{F_\Phi}+\frac{1}{F_\phi}\right)
\ln(m_P/m_{3/2})
&,
\label{Ntot}
\end{eqnarray}
where \mbox{$F_\Phi\geq\frac{3}{2}$} and 
\mbox{$N_\Phi\equiv\max\{N_{\rm FR}, N_{\rm lock}\}$} corresponds to the first 
stage of inflation. It is $N_{\rm tot}$ that has to be compared to the 
necessary e-foldings for the cosmological scales.

\section{The necessary e-foldings}

The inflationary period has to be sufficiently long to encompass the 
cosmological scales. This results in the following lower bound on the total 
number of e-foldings of inflation~\cite{book}:
\begin{eqnarray}
N_C\;= &
\!\!72-\ln\left(m_P/V_{\rm inf}^{1/4}\right)-
\frac{1}{3}\,\ln\left(V_{\rm inf}^{1/4}/T_{\rm reh}\right)
&,
\label{N}
\end{eqnarray}
where \mbox{$T_{\rm reh}\sim\sqrt{\Gamma m_P}$}, is the reheat temperature, 
\mbox{$\Gamma\simeq g^2m_\phi$} is the decay rate for the inflaton field $\phi$
(corresponding to the last stage of inflation)
and $g$ is the coupling of $\phi$ to the decay products. If the coupling of 
the field to other particles is extremely weak then $\phi$ decays 
gravitationally, in which case \mbox{$\Gamma\sim m_\phi^3/m_P^2$}. Thus, the 
effective range for $g$ is: \mbox{$m_{3/2}/m_P\leq g\leq 1$}.

Using that \mbox{$V_{\rm inf}\sim M_S^4$}, 
we find:
\mbox{$N_C=54-\frac{1}{3}\ln g
$}.
Demanding that \mbox{$N_{\rm tot}>N_C$} provides an upper bound on $m_\phi$, 
which is more stringent the smaller $N_\Phi$ is. Hence, the tightest bound 
corresponds to \mbox{$N_\Phi=N_{\rm lock}$}. After a little algebra we obtain 
the bound
%
%
\begin{eqnarray}
m_\phi/H_{\rm inf}\;< &
\!\!\frac{3}{2}\,
\Big\{\Big[\Big(
\frac{108+\ln\sqrt{g}
}{\ln(m_P/m_{3/2})}-\frac{7}{4}
\Big)^{-1}+1\Big]^2-1\Big\}^{1/2}.
 & \label{bound}
\end{eqnarray}
Considering the range of $g$ it is easy to check that the above bound 
interpolates between 2 and~3. Thus, it is possible to satisfy the cosmological 
observations with \mbox{$m_\Phi\sim m_\phi\sim m_{3/2}\sim H_{\rm inf}$}.
Hence, {\em the combination of locked and fast-roll inflation is capable of 
providing enough e-foldings to encompass the cosmological scales without the 
use of any flat direction}. 
If the mass of $\Phi$ is below $\frac{3}{2}H_{\rm inf}$ then $N_\Phi$ is given 
by \mbox{$N_{\rm FR}>N_{\rm lock}$}. In this case the bound on $m_\phi$ is 
further relaxed ($N_\phi$ does not need to be as large). Hence, {\em 
regardless of $m_\Phi$, 
the required e-foldings 
corresponding to the cosmological scales, 
can be obtained
only with a 
mild upper bound on 
$m_\phi$}.

\section{Discussion and conclusions}

Using natural values for the parameters 
and a generic, hybrid-type potential we showed that 
moduli fields, corresponding to flat directions of supersymmetry, whose 
flatness is lifted by supergravity corrections, can naturally generate enough 
e-foldings of inflation to solve the horizon and the flatness problems with 
only a mild upper bound on the tachyonic mass of the inflaton
and without employing slow roll at all. That way inflation can escape from the 
famous $\eta$-problem.

Structure formation, in our model, is due to the existence of a curvaton field,
which is not linked to the moduli inflatons. The curvaton {\em must} be a flat 
direction. Being unrelated to inflationary dynamics, the curvaton can be 
protected by a symmetry (other than supersymmetry), which may even be exact 
during inflation (e.g. a global U(1) for a PNGB curvaton
\cite{pngb}). 

The tachyonic fluctuations at the phase transition which terminates the first
stage of inflation can generate 
primordial black holes with mass comparable to the mass of the horizon volume 
at the time of their creation \cite{pbh}. 
Hence, the earlier they form the smaller they are and the sooner they 
evaporate. Therefore, to protect nucleosynthesis, $V_{\rm inf}$ has to be 
bounded from below as~\cite{Vbound} 
\mbox{$V_{\rm inf}^{1/4}\gsim 10^{11}$GeV}, which our model marginally 
satisfies. Problems may also arise from the possible formation of topological 
defects at the phase transition, depending on their stability.
Nevertheless, both black holes and defects can be avoided provided that
\mbox{$\phi_0>m_\phi$}.

\end{document}